\documentclass{sf2a-conf2023}
\usepackage{graphicx}
\usepackage{hyperref}
\usepackage[]{natbib}  
\usepackage{epstopdf}
\usepackage{amsmath, amssymb, amsfonts, graphicx}
\usepackage{mathtools}

\def\BibTeX{{\rm B\kern-.05em{\sc i\kern-.025em b}\kern-.08em
    T\kern-.1667em\lower.7ex\hbox{E}\kern-.125emX}}
\bibpunct{(}{)}{;}{a}{}{,}  %%%%%%%%%%%%%  A&A bibliography style
\begin{document}

\TitreGlobal{SF2A 2023}

%%-----------------------------------------------------------------
%%      the top matter
%%

\title{Hydrodynamical modelling of tidal dissipation in gas giant planets at the time of space missions}

\runningtitle{Hydrodynamical modelling of tidal dissipation in gas giant planets at the time of space missions}
\author{H. Dhouib}\address{Université Paris Cité, Université Paris-Saclay, CEA, CNRS, AIM,  F-91191, Gif-sur-Yvette, France}
\author{C. Baruteau}\address{IRAP, Université de Toulouse, CNRS, UPS, F-31400 Toulouse, France}
\author{S. Mathis}\address{Université Paris-Saclay, Université Paris Cité, CEA, CNRS, AIM,  F-91191, Gif-sur-Yvette, France}    
\author{F. Debras$^2$}   
\author{A. Astoul}\address{Department of Applied Mathematics, School of Mathematics, University of Leeds, Leeds LS2 9JT, UK}    
\author{M. Rieutord$^2$}

%% IF Author3 has the same affiliation than Author1:
%\author{C.\,E. Author3$^1$}

%% IF Author3 has its own affiliation:
%\author{C.\,E. Author3}\address{Dept. of Chess, University of Games, 35101 Las Vegas, Monaco} 

%% IF Author3 has two affiliations, the one of Author1 and a second one:
%\author{C.\,E. Author3$^{1,}$}\address{Dept. of Chess, University of Games, 35101 Las Vegas, Monaco} 

%% Keep this line, even if the page will be settled afterwards.
\setcounter{page}{237}

%%-----------------------------------------------------------------

\maketitle

%%-----------------------------------------------------------------
%%        The abstract
%% 
%%  Warning!  within the abstract:
%%  - do not use macros. 
%%  - do not use commands like: \cite, \citet, \citep ... etc.

\begin{abstract}
Gas giant planets are differentially rotating magnetic objects that have strong and complex interactions with their environment. In our Solar system, they interact with their numerous moons while exoplanets with very short orbital periods (hot Jupiters), interact with their host star. The dissipation of waves excited by tidal forces in their interiors shapes the orbital architecture and the rotational dynamics of these systems.
Recently, astrometric observations of Jupiter and Saturn systems have challenged our understanding of their formation and evolution, with stronger tidal dissipation in these planets than previously predicted, in contrast to what appears to be weaker in gas giant exoplanets. These new constraints are motivating the development of realistic models of tidal dissipation inside these planets. At the same time, the Juno and Cassini space missions have revolutionised our knowledge of the interiors of Jupiter and Saturn, whose structure is a combination of stably stratified zones and convective regions.
In this work, we present results of hydrodynamical calculations modelling tidal waves and their dissipation in Jupiter, taking for the first time the latest, state-of-the-art interior model of the planet. We performed 2D numerical simulations of linear tidal gravito-inertial waves that propagate and dissipate within Jupiter interior by taking into account viscous, thermal and chemical diffusions. This new model allows us to explore the properties of the dissipation and the associated tidal torque as a function of all the key hydrodynamical and structural parameters.
\end{abstract}
%% Insert the keywords (to appear in the ADS indexing)
%% Keywords must be separated by a comma

%%-----------------------------------------------------------------

\section{Introduction}
Tides dissipate energy through various mechanisms, such as turbulent friction and heat diffusion \citep[e.g.][]{Ogilvie2014, Mathis2019}. This dissipation impacts the evolution of planet-moon systems. Jupiter and Saturn display unexpectedly strong tidal dissipation, driving rapid orbital migration revealed by precise astrometric data \citep{Lainey2009, Lainey2012, Lainey2017, Lainey2020}. For example, \cite{Lainey2009} determined a tidal dissipation rate of $k_{2 2}/Q = \left(1.1 \pm 0.2 \right)\times 10^{-5}$ \footnote{$k_{\ell m}$ is the tidal Love numbers which quantitatively characterises the planet's adiabatic hydrostatic response to the $(\ell, m)$ component of the tidal forcing, where $\ell$ and $m$ are the latitudinal degree and azimuthal order of the corresponding spherical harmonics. $Q$ is the tidal quality factor which evaluates the ratio
between the maximum energy stored in the tidal distortion and the energy dissipated during an orbital period.} for Io's asynchronous tide. This is one order of magnitude stronger than previous theoretical predictions based on moon formation scenarios \citep{Goldreich1966}.
In addition, the space missions Juno and the grand finale of the Cassini mission have completely changed our vision of the interiors of Jupiter and Saturn \citep{Wahl2017, Guillot2018, Galanti2019}. They revealed that these planets are structured by a central stably stratified core, a convective metallic shell, a potential intermediate stable layer in the case of Jupiter, and an outer differentially rotating molecular convective envelope.
These new constraints motivate the development of realistic models of tidal dissipation inside these planets. That is why we develop a method to compute the dissipation of the dynamical tidal response of a  self-gravitating, rotating fluid body composed of alternating convective layers and stably stratified layers \citep[i.e. tidally-excited gravito-inertial waves; see also][]{Lin2023, Dewberry2023} and which takes into account the viscous, thermal and chemical dissipation processes. We focus here on the latest Jupiter interior models built by \cite{Debras2019}, which matches constraints obtained by the Juno space mission.

\section{Tidally forced waves in gas giant planet interiors}
We investigate the linear excitation of (gravito-)inertial waves induced by external tidal forces. The dynamics of these waves are governed by Coriolis acceleration and buoyancy in stable layers and they undergo dissipative processes (assumed to be uniform here), namely viscosity ($\nu$), thermal diffusion ($\kappa$), and molecular diffusion ($\mathrm{D}_\mu$).
First, we linearise the hydrodynamic system around the hydrostatic steady-state. Each scalar field $X\coloneqq\{\rho, \Phi, T, \mu, P\}$\footnote{$\rho$, $\Phi$, $T$, $\mu$ and $P$ are the density, the gravitational potential, the temperature, the mean molecular weight, and the pressure, respectively.} is expanded as the sum of its hydrostatic value $X_0$ and of the Eulerian perturbations associated with the tides $X^\prime$:
$X(r,\theta,\varphi, t)=X_0(r)+{X}^\prime(r,\theta,\varphi, t)$
and the  velocity  field, $\vec{V}$, is expanded as  the sum of the large-scale azimuthal velocity associated with the uniform rotation  $\Omega$\footnote{As a first step we neglect differential rotation, since Jupiter's relative differential rotation is 4\% \citep{Guillot2018}.} and of the wave velocity $\vec{v}$:
    $\vec{V}(r,\theta,\varphi, t)= r \sin{\theta} \, \Omega \,  \vec{e}_{\varphi} + \vec{v}(r,\theta,\varphi, t)$, where $t$ is time and $(r,\theta,\varphi)$ are the usual spherical coordinates with their associated unit vector basis $(\vec{e}_r, \vec{e}_\theta, \vec{e}_\varphi)$.
Afterwards, we decompose the fluctuations associated with the tides into non-wavelike and wavelike parts: $Y = Y^{\rm nw} + Y^{\rm w}$,
where $Y\coloneqq\{v_r, v_\theta, v_\varphi, X^{\prime}\}$, $ Y^{\rm nw}$ is the non-wavelike (equilibrium) tide that satisfies the hydrostatic equilibrium, and $Y^{\rm w}$ the wavelike (dynamical) tide that describes the propagation of tidal waves.
The forcing term ($f^{\ell,m}$) arises when solving the wavelike tide as a residual force coming from the subtraction of the hydrostatic balance verified by the non-wavelike tide from the complete momentum equation \citep{Ogilvie2014}.
This force encompasses the acceleration of the non-wavelike tide and the Coriolis acceleration applied to it, which forces the (gravito-)inertial tidal waves.
Finally, by assuming Boussinesq and Cowling approximations as a first step and by expanding on (vectorial) spherical harmonics \citep{Rieutord1987} the velocity field $(u^{m}_{\ell}, v^{m}_{\ell}, w^{m}_{\ell})$, temperature ($t^{m}_{\ell} $), molecular weight ($\mu^{m}_{\ell} $), reduced pressure ($p^{m}_{\ell} $) and the forcing vector ($f^{\ell,m}_{\rm R}, f^{\ell,m}_{\rm S}, f^{\ell,m}_{\rm T}$), we can write the dimensionless system that describe the wavelike tides as :
\begin{equation}
    \mathrm{d}_{r}u^{m}_{\ell} + \frac{2}{r}  u^{m}_{\ell} - \ell(\ell+1) \frac{v^{m}_{\ell}}{r} = 0,\label{eq:solve1}
\end{equation}
\begin{multline}
       \mathrm{E}\Delta_{\ell} u^{m}_{\ell}-\left(\frac{2\mathrm{E}}{r^{2}} - i \tilde{\omega}\right) u^{m}_{\ell} + \left(i m+ \frac{2\mathrm{E}}{r^{2}} \ell(\ell+1)\right) v^{m}_{\ell} - \\ \beta_{\ell-1}^{\ell} w_{\ell-1}^m - \beta_{\ell+1}^{\ell} w_{\ell+1}^m  -\mathrm{d}_r p^{m}_{\ell} + \frac{\delta g_0^*}{T_0^*} t^{m}_{\ell}  - \frac{\phi  g_0^*}{\mu_0^*} \mu^{m}_{\ell}  = - f^{\ell,m}_{\rm R},
\end{multline}
\begin{equation}
       \mathrm{E} \Delta_{\ell} v^{m}_{\ell}+ \left(i\tilde{\omega} + \frac{i m}{\ell(\ell+1)} \right) v^{m}_{\ell} +\left(\frac{2\mathrm{E}}{r^{2}} + \frac{i m}{\ell(\ell+1)}\right) u^{m}_{\ell}
         -  \gamma_{\ell-1}^{\ell} w_{\ell-1}^m - \gamma_{\ell+1}^{\ell} w_{\ell+1}^m  -\frac{p^{m}_{\ell}}{r} =-f^{\ell,m}_{\rm S},
\end{equation}
\begin{equation}
       \mathrm{E} \Delta_{\ell} w^{m}_{\ell} +\left(i\tilde{\omega} + \frac{i m}{\ell(\ell+1)} \right)w^{m}_{\ell} + \gamma_{\ell-1}^{\ell} v_{\ell-1}^m + \gamma_{\ell+1}^{\ell} v_{\ell+1}^m -  
       \frac{\alpha_{\ell-1}^{\ell}}{\ell} u_{\ell-1}^m + \frac{\alpha_{\ell+1}^{\ell}}{\ell+1} u_{\ell+1}^m = - f^{\ell,m}_{\rm T},
\end{equation}

\begin{equation}
     \frac{\mathrm{E}}{\mathrm{Pr}}\Delta_\ell t^{m}_{\ell}  - u^{m}_{\ell}  \frac{T_0^*{N_{\rm t}^*}^{2}}{g_0^* \delta} + i\tilde{\omega} t^{m}_{\ell}=0, \quad      \frac{\mathrm{E}}{\mathrm{Sc}}\Delta_\ell \mu^{m}_{\ell}  + u^{m}_{\ell} \frac{\mu_0^*{N_\mu^*}^{2}}{g_0^* \phi} + i\tilde{\omega} \mu^{m}_{\ell}=0,\label{eq:solvelast}
\end{equation}
where $\tilde{\omega}$ is the normalised tidal forcing frequency, $g_0^*$ is the normalised gravitational acceleration, ${N_{\rm t}^*}^{2}=N_{\rm t}^2/4 \Omega^{2}$ is the normalised Brunt–Väisälä frequency linked to the thermal stratification squared, ${N_\mu^*}^{2}=N_\mu^2/4 \Omega^{2}$ is the normalised Brunt–Väisälä frequency linked to the chemical stratification squared, and 
\begin{gather}
    \Delta_{\ell}=\mathrm{d}_{r^2}+\frac{2}{r} \mathrm{d}_{r}-\frac{\ell(\ell+1)}{r^{2}},\quad
    \alpha_{\ell-1}^{\ell}=\alpha_{\ell}^{\ell-1}=\sqrt{\frac{\ell^{2}-m^{2}}{(2 \ell-1)(2 \ell+1)}}, \quad \beta_{\ell-1}^{\ell}=(\ell-1) \alpha_{\ell-1}^{\ell}, \\
     \beta_{\ell+1}^{\ell}=-(\ell+2) \alpha_{\ell+1}^{\ell},\quad
    \gamma_{\ell-1}^{\ell}=\frac{\ell-1}{\ell} \alpha_{\ell-1}^{\ell}, \;
    \gamma_{\ell+1}^{\ell}=\frac{\ell+2}{\ell+1} \alpha_{\ell+1}^{\ell},\; 
    \delta\coloneqq-\left(\frac{\partial \ln \rho}{\partial \ln T}\right)_{P,\,\mu}, \; \phi\coloneqq\left(\frac{\partial \ln \rho}{\partial \ln \mu}\right)_{P,\,T}.
\end{gather}
We define also the following dimensionless numbers : the Prandtl number $\mathrm{Pr}=\nu/\kappa$, the Schmidt number $\mathrm{Sc}=\nu/D_\mu$ and the Ekman number $\mathrm{E}=\nu/2 \Omega R^{2}$\footnote{R is Jupiter's radius.} (the ratio between the viscous force and the Coriolis force).
\section{Jupiter's interior model matching the constraints provided by Juno}
\begin{figure}[h!]
    \centering
    \includegraphics[width=0.48\textwidth,clip]{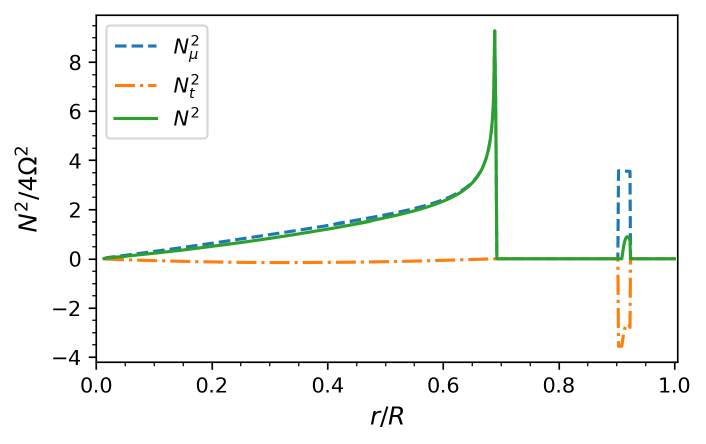}      
    \includegraphics[width=0.35\textwidth,clip]{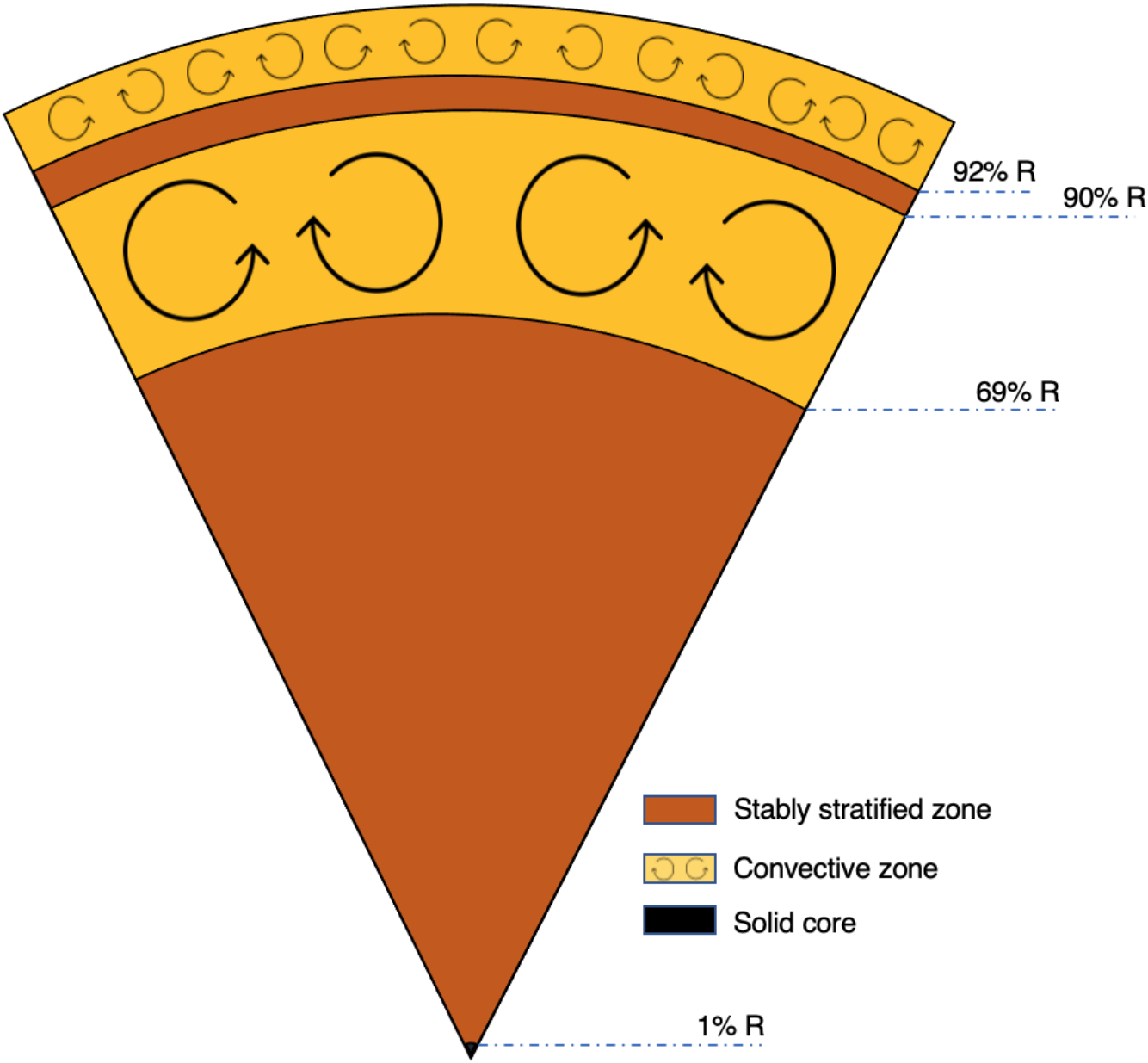}      
    \caption{{\bf Left:} Radial profiles of the normalised compositional ($N_\mu^2$), thermal ($N_{\rm t}^2$), and total ($N^2$) buoyancy frequencies squared of Jupiter’s interior used in this study. {\bf Right:} Schematic of this model. }
    \label{dhouib:fig1}
\end{figure}

The internal structure model that we consider here is computed by \cite{Debras2019, Chabrier2021} to reproduce Jupiter's multipolar  gravitational moments as measured by Juno. 
As illustrated in Fig.\,\ref{dhouib:fig1}, starting from its surface and moving towards the core, Jupiter is thought to exhibit the following layers : a gaseous convective envelope, a transitional stably stratified  zone considered to be potentially semi-convective, an internal convective zone composed of metallic hydrogen and helium, a stably stratified zone located closer to the core which may exhibit double diffusion convection or a diluted core structure due to stabilising composition gradients, and a potential unstable solid core of size $1.4\%$ of the radius made up of rock or ice \footnote{Note that as a first step, we neglected in this study the differential rotation in the outer convective region and the magnetic field in the internal one.}. 
\section{Dissipation spectra and imaginary part of the Love number}
\begin{figure}[ht!]
    \centering
    \includegraphics[width=0.72\textwidth,clip]{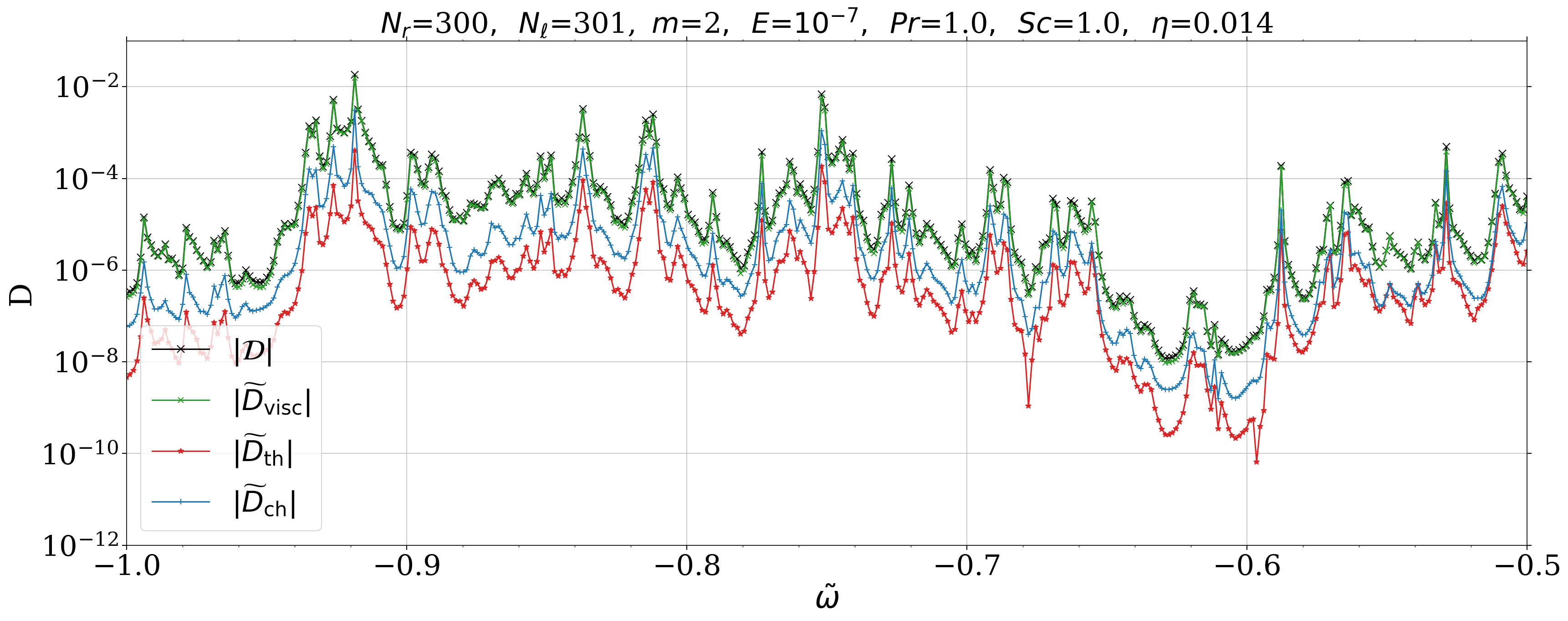}
    \includegraphics[width=0.72\textwidth,clip]{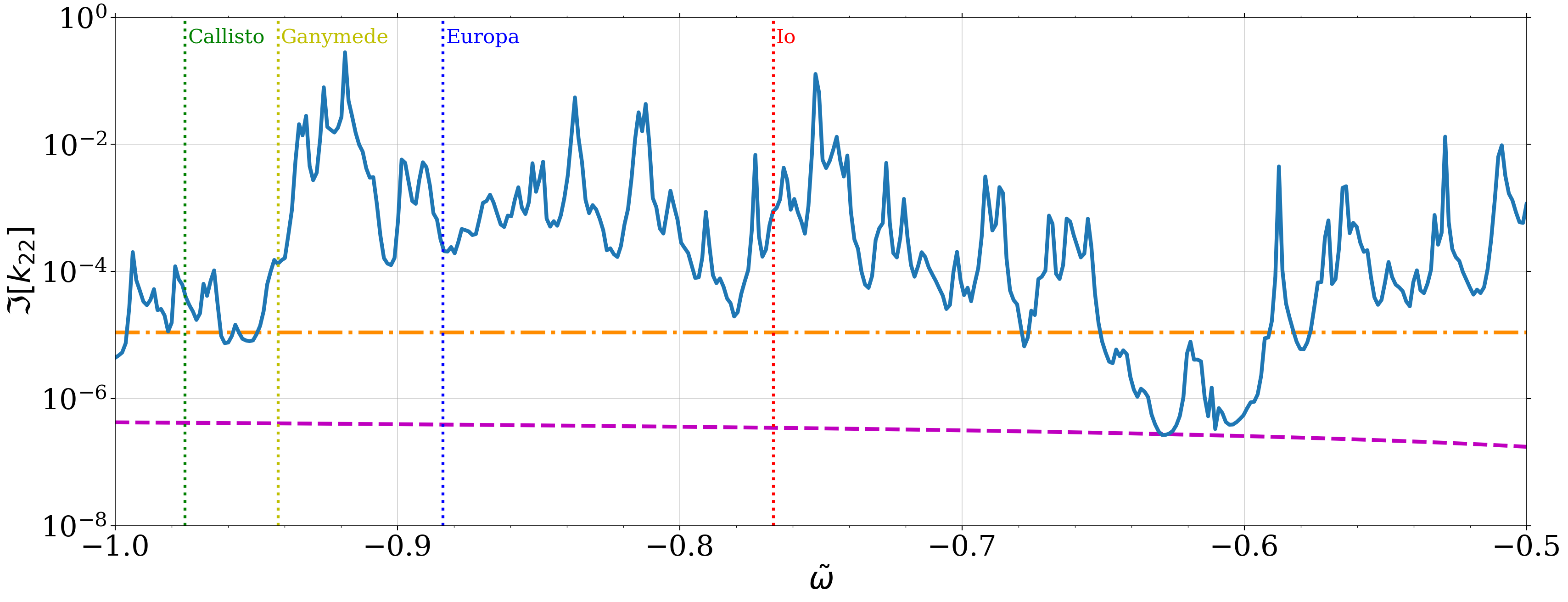}      
    \caption{Dissipations (Top) and imaginary part of the Love number (Bottom) as a function of the tidal frequency for $m=2$, $\mathrm{Pr}=\mathrm{Sc}=1$,  $\mathrm{E}=10^{-7}$ and $(N_r,\,N_\ell)=(300,\,301$). The magenta dashed line indicates the values of these quantities in the case of a purely convective interior. Vertical dotted lines indicate the tidal frequencies for the four Galilean Moons of Jupiter. The dash-dotted orange line marks the observed value due to Io \citep{Lainey2009}.}
    \label{author1:fig2}
\end{figure}

We solve the system \eqref{eq:solve1}-\eqref{eq:solvelast} numerically using the 2D pseudo-spectral linear code LSB \citep[Linear Solver Builder,][]{Valdettaro2007}.
These equations are discretised in the radial direction on the Gauss-Lobatto collocation nodes associated with the Chebyshev polynomials. They are truncated to order $N_r$ for the Chebyshev basis and to order $N_\ell$ for the spherical harmonics basis. We specifically investigate the quadrupolar tidal components ($\ell=m=2$).
The top panel of the Fig.\,\ref{author1:fig2} shows the viscous ($\widetilde{D}_{\mathrm{visc}}$), thermal ($\widetilde{D}_{\mathrm{th}}$), molecular ($\widetilde{D}_{\mathrm{ch}}$) and total ($\mathcal{D}$) dissipation rates integrated over the volume as a function of the normalised forcing frequency ($\tilde{\omega}$) for $\mathrm{E}=10^{-7}$ and $\mathrm{Pr}=\mathrm{Sc}=1$. We observe a significant frequency dependence, indicating a strong relationship between dissipation and forcing frequency. Moreover, for our set of $\mathrm{E}$, $\mathrm{Sc}$ and $\mathrm{Pr}$, the dominant mechanism contributing to dissipation is viscosity, surpassing both thermal and chemical dissipations in magnitude. We also represent the dissipation spectra for the standard vision of Jupiter's interior before Juno's results, where there is a single purely convective zone extending from $r=\eta=0.014$ to $r=1$. We can see that the spectra in this case exhibit a smooth profile with weaker magnitude, devoid of any pronounced peaks at specific frequencies.
The bottom panel of Fig.\,\ref{author1:fig2} shows significant discrepancy between computed values of the imaginary part of the Love number due to Io and the observed ones, differing by roughly two orders of magnitude. Consequently, our calculations tend to overestimate the amplitude of tidal dissipation. Conversely, when examining the purely convective model, we observe an underestimation of tidal dissipation by approximately one and a half orders of magnitude. This shows the key role played by stably stratified layers in controlling the strength of the dissipation.

\section{Conclusions}
We examine the dissipation of dynamical tides in the latest Jupiter interior multi-layer model with alternating convective and stably stratified regions. We take into consideration various types of dissipations such as turbulent viscosity, thermal dissipation, and molecular diffusivity. This enables a more comprehensive and realistic representation of the physical processes occurring within giant gas planets’ interiors. We find that the presence of stably stratified regions plays a significant role in explaining the strong dissipation observed in Jupiter when compared to the case of a sole convective envelope. It is important to note that our model is not limited to Jupiter but can also be applied to other giant planets such as Saturn, as well as exoplanets.\footnote{Acknowledgments: this work has been supported by PNP (CNRS/INSU) and the PLATO CNES grant at CEA/DAp and by the Leverhulme Trust for the award of an Early Career Fellowship to AA (ECF-2022-362).}

\bibliographystyle{aa}  % A&A bibliography style file (aa.bst)
\bibliography{Dhouib} % your references in file: Yourfile.bib

\end{document}